\documentclass{iaus}
\usepackage{graphicx}

\title[JD 11.~~Extending the RP survey to the outer LMC] %% give here short title %%
{Extending the RP survey to the outer LMC}

\author[Warren A. Reid \& Quentin Parker]   %% give here short author list %%
{Warren Reid$^1\,^2$
 \and Quentin A. Parker$^1\,^2\,^3$}

\affiliation{$^1$Department of Physics and Astronomy, Macquarie University, Sydney, NSW 2109, Australia  \\
$^2$Macquarie University Research Centre in Astronomy, Astrophysics \& Astrophotonics \\email: {\tt warren.reid@mq.edu.au} \\[\affilskip]
$^3$Australian Astronomical Observatory, PO Box 296, Epping, NSW 1710, Australia \\email: {\tt quentin.parker@mq.edu.au}}

\pubyear{2011}
\volume{283}  %% insert here IAU Symposium No.
\pagerange{119--126}
\jname{Planetary Nebulae an Eye to the Future}
\editors{A.C. Editor, B.D. Editor \& C.E. Editor, eds.}
\begin{document}

\maketitle

\begin{abstract}
We are extending our search for faint PNe in the LMC to include the outer 56 deg$^{2}$ area not covered in the original UKST survey of the central 25deg$^{2}$ region. Candidate PNe have been selected using the Magellanic Cloud Emission Line Survey (MCELS) and the first round of observations has yielded 93 new LMC PNe while confirming the 102 previously known PNe in the outer LMC. We plan to continue our spectroscopic object identification program until we cover all our remaining candidates in the survey area. These observations, providing medium and high resolution spectra from $\lambda$3650\AA~to $\lambda$6900\AA~will additionally be used to measure fluxes for a series of research projects including luminosity functions, abundances and LMC kinematics.
\end{abstract}
\firstsection % if your document starts with a section,
\section{Introduction}
\begin{figure}[h]
% \vspace*{-2.0 cm}
\begin{center}
 \includegraphics[width=3.4in]{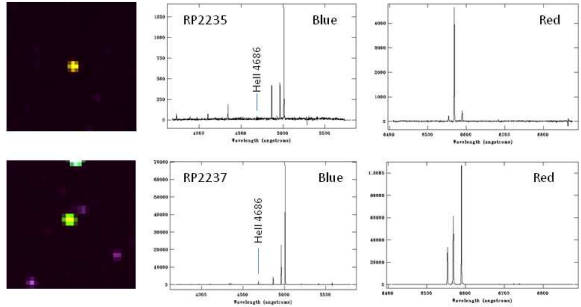}
% \vspace*{-1.0 cm}
 \caption{Left: the candidate image from MCELS (1$\times$1 arcmin). Middle: the confirmatory blue 580V spectrum from AAOmega. Right: the confirmatory red 1000R spectrum from AAOmega. }
   \label{fig1}
\end{center}
\end{figure}

In 2006 Reid and Parker added 460 planetary nebulae (PNe) to the 169 previously known across the central 25deg$^{2}$ of the LMC. These candidates were assigned a probability rating which relied on spectroscopic and optical confirmation. With the assistance of increased high resolution spectroscopy and NIR imaging from Spitzer Space Telescope SAGE data (see \cite[Meixner et al. 2006]{Meixner2006}) we have been able to either confirm or re-classifying most of the lower classed ``possible PNe''. The resulting large number of PNe now known in the central LMC (\cite[Reid \& Parker, 2006a,b]{Reid2006b}) have yielded significant advances in our knowledge of the central LMC's kinematical sub-structure (rotation, inclinations, transverse velocity) and raised interesting questions regarding the kinematical structure of the outer regions. Our access to the MCELS survey (\cite[Smith et al. 1998]{Smith1998}) provides the opportunity to achieve equivalent results across a much larger 84deg$^{2}$ of the LMC, critical for a complete determination of kinematics, abundance gradients and central star properties (eg. \cite[Reid \& Parker, 2010b]{Reid2010b}) . We have already used AAOmega to spectroscopically verify 93 new PNe to add to the 102 previously known in the outer LMC (eg. Fig.\,\ref{fig1}). New PN radial velocities are being compared to other tracers and the HI gas disk. These are being added to existing kinematic data to create gradients and verify models. With a near complete LMC PNe census to V=22, an unbiased LMC PNe luminosity function (PNLF) is being built in order to identify any population sub-trends or `dips' while providing an accurate bright-end cut off which is used as a standard candle.
\section{The Planetary Nebula Luminosity Function}
The confirmation of the previously known and faint new PNe candidates in the outer LMC to V=21 will provide the most accurate PNLF across a whole galaxy ever achieved. Relatively few  faint PNe have been observed in previous surveys and the faintest three magnitudes of the LMC PNLF were very poorly determined. We have already overcome this limitation for the main LMC bar (\cite[Reid \& Parker 2010a]{Reid2010a}). A PNLF across a whole galaxy can accurately predict the number of stars in the PN evolutionary stage and in each luminosity bin compared to the luminosity and mass of the whole galaxy. This is not feasibly attainable for any galaxy other than the LMC and SMC, as dust obscures too much of our own Galaxy and other galaxies are too distant to identify faint PNe. Our new LMC PNLF will include a far more complete sample in terms of depth, coverage and PN evolutionary state covering an unprecedented 10 magnitude range.
\section{A complete LMC kinematic map using PNe}
Following positive identification of all outer LMC PN candidates, their positions and subsequent measured spectroscopic velocities will be used to map the kinematic structure of the LMC to investigate warps, velocity structures and system rotation. Our previous 2dF data provided (~5km/s) velocities for PNe in the LMC's central bar which enabled \cite[Reid \& Parker, (2006b)]{Reid2006b} to investigate the kinematics of the PNe population and their interaction with the tidal forces and the HI disk on the main bar. \cite[Reid \& Parker, (2006b)]{Reid2006b} discovered slow, solid body rotation of the PN population in together with the HI disk on the main bar revealing a major disruption to PNe relative to the HI disk either side of the main bar. This becomes clearly evident when measuring the transverse velocity of the central region. \cite[Reid \& Parker, (2006b)]{Reid2006b} also found an elliptical shift in the PN kinematic line of nodes which is offset to the north ($\Theta$ = 184$^{\circ}$). This represents an increasing displacement along this line in the central 25deg$^{2}$ region extending outwards from the main bar. Our latest observations of the outer LMC are allowing us to discover the extent of this warping of the old population disk along the line of nodes as it extends outwards. Confirmation of this extended structure, using PNe as intermediate stellar tracers, will impact LMC evolutionary theories since it indicates strong tidal interaction affecting old populations more than young. Using the velocity dispersion profile of PNe across the whole LMC, we will be able to determine the tidal radius and estimate the LMC mass.

\end{document}